\documentclass[letter]{aa}
\usepackage{graphicx}
\usepackage{epsfig}
\usepackage{latexsym}
\usepackage{txfonts}
\usepackage{natbib}
\begin{document}
\renewcommand{\labelitemi}{-}
\title{Do solar decimetric spikes originate in coronal X-ray sources?}
\author{Marina Battaglia
  \and Arnold O. Benz }
\institute{Institute of Astronomy, ETH Zurich, 8092 Zurich, Switzerland}
\date{Received /Accepted}

\abstract
{In the standard solar flare scenario, a large number of particles are accelerated in the corona. Nonthermal electrons emit both X-rays and radio waves. Thus, correlated signatures of the acceleration process are predicted at both wavelengths, coinciding either close to the footpoints of a magnetic loop or near the coronal X-ray source. }
{We attempt to study the spatial connection between coronal X-ray emission and decimetric radio spikes to determine the site and geometry of the acceleration process.}
{The positions of radio-spike sources and coronal X-ray sources are determined and analyzed in a well-observed limb event. Radio spikes are identified in observations from the Phoenix-2 spectrometer. Data from the Nan\c{c}ay radioheliograph are used to determine the position of the radio spikes. RHESSI images in soft and hard X-ray wavelengths are used to determine the X-ray flare geometry. Those observations are complemented by images from GOES/SXI.  }
{We find that the radio emission originates at altitudes much higher than the coronal X-ray source, having an offset from the coronal X-ray source amounting  to 90\arcsec\ and to 113\arcsec\ and 131\arcsec\ from the two footpoints, averaged over time and frequency.}
{Decimetric spikes do not originate from coronal X-ray flare sources contrary to previous expectations. However, the observations suggest a causal link between the coronal X-ray source, related to the major energy release site, and simultaneous activity in the higher corona.}

\keywords{Sun: flares -- Sun: X-rays, gamma-rays -- Sun: radio radiation -- Acceleration of particles}
\titlerunning{Do decimetric spikes originate from coronal X-ray sources?}
\authorrunning{Marina Battaglia \& Arnold O. Benz}

\maketitle


\section{Introduction} \label{Introduction}
Signatures of particles accelerated in solar flares are generally found in hard X-rays as well as radio wavelengths.
Hard X-ray sources at altitudes of several $10^4$ km  in the corona have been suspected to be possible sites of particle acceleration. Their existence was first proposed by \citet{Fr71} and confirmed by the discovery of a hard X-ray source above the top of a magnetic loop by \citet{Ma94}. Statistical studies using Yohkoh data \citep{To01} and RHESSI \citep{Kr08} confirmed the existence of a hard X-ray component in 90\% of the analyzed coronal sources. However, Masuda-type events with a significant displacement of the coronal hard X-ray component from the soft X-ray component are rare. A comprehensive review of coronal hard X-ray sources can be found in \citet{Krrev08}.
\begin{figure*}[t!h]
\centering
\includegraphics[width=13.5cm]{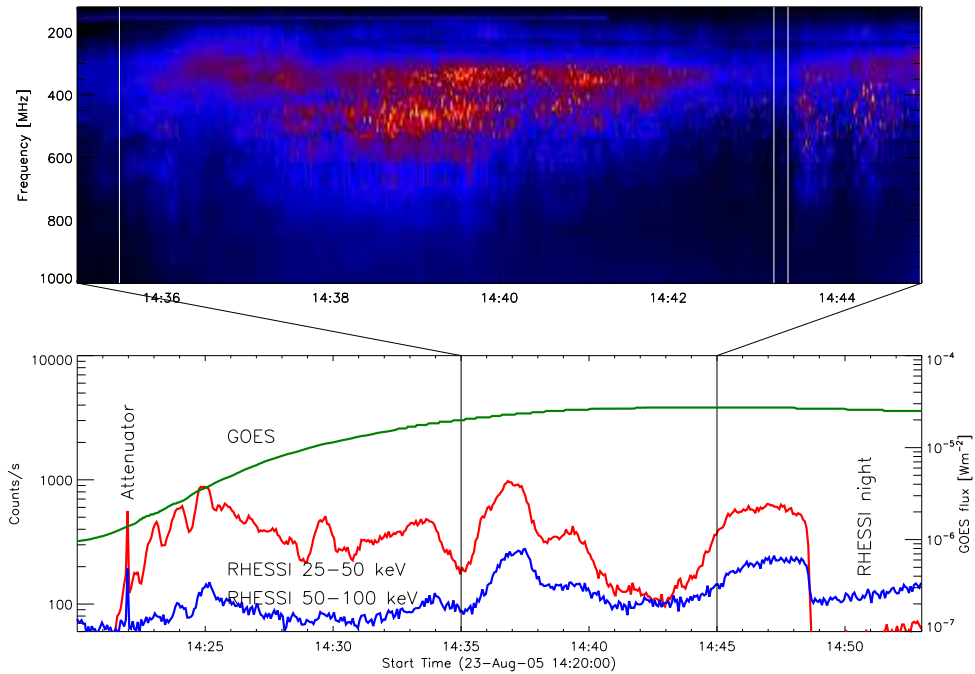}
\caption {\textit{Top}: Phoenix-2 spectrogram. \textit{Bottom}: RHESSI light-curves at 25-50 keV and 50-100 keV. GOES light-curve in the 1-8 \AA\ band. Time intervals containing the spikes are indicated by white vertical lines in the spectrogram.} 
\label{spectrogram}
\end{figure*}

Several observations support the scenario that the coronal X-ray source is closely linked to both the acceleration site and mechanism. \citet{Ba06} analyzed five RHESSI events using imaging spectroscopy techniques to separate the coronal emission from the footpoint emission. They found that the time evolution of the coronal-source spectra follows the so-called soft-hard-soft pattern. This spectral property has been studied since its discovery in full sun spectra \citep[e.g.,][]{Pa69,Gr04}, corresponding mostly to emission from the footpoints. Since it is observed in the coronal source, this phenomenon is probably not a transport effect. It can indeed be reproduced by acceleration theories \citep[e.g.,][]{Gr06,By09}, suggesting acceleration in the coronal source. Further evidence of a close connection between the coronal source and acceleration is provided by \citet{Kr08} who identified rapid time-variations in the hard X-ray component of coronal sources indicative of nonthermal electrons not yet affected by propagation. 

The flare radio emission at decimeter wavelengths also originates from nonthermal electrons. Thus, a close association with hard X-rays may be expected. The most powerful emission is coherently radiated by plasma waves or instabilities (masers).  Among the various coherent radio phenomena at decimeter wavelengths, millisecond spikes have the highest association rate with hard X-rays \citep[e.g.,][]{Be86, Gu91, Asch92}. Initially, decimetric spikes attracted much attention because of their discovery during the rise phase of centimeter (gyro-synchrotron) emission of solar flares \citep{Sl78}. Thus, among the coherent radio phenomena, they are expected to have the closest relation to the energy release and subsequent particle acceleration. An observational overview of the spike phenomenon is given in \citet{Be86}. 

The spatial relation between hard X-rays and spikes is more controversial than their temporal association. Two different spike locations are predicted by the theories of spike emission. In theories proposing electron cyclotron maser-emission caused by loss-cone instability of trapped electrons \citep[e.g.,][]{Ho80,Mel82,Fle94}, the emission would be expected to originate in regions close to the footpoints of a magnetic loop. In the other theories, spike sources are proposed to result from waves produced in the acceleration process and would therefore be expected to originate from the location of the acceleration \citep{Ta90, Gu93}. In a flare scenario as described above where the acceleration is thought to be strongly related to the coronal X-ray source, the location of the spikes  would then be expected to be at or close to the position of coronal hard X-ray sources. \citet{Be02rad} studied a number of solar flares with associated hard X-ray sources and spikes, finding the location of the spikes to be offset by up to 400'' from the flaring site as observed in hard X-ray footpoints, soft X-rays, and EUV. Since their events were observed on the solar disk, the complete geometry of the events could not be determined. \citet{Kh06} analyzed a limb event, finding that the spike emission originates high in the corona. No coronal hard X-ray source was observed in the events studied to date. 

Where do spikes occur relative to coronal hard X-ray flare emission? RHESSI \citep{Li02} now provides the possibility of studying the X-ray morphology of flares, including the hard X-ray component of coronal sources. Here, we study the spatial relations between the X-ray sources, both soft and hard, and the position of the radio emission in a spike event. We find that the locations of the spikes do not coincide with the locations of either the footpoints or the coronal X-ray source, but seem to be completely disconnected from the flaring site, as is clearly seen in X-rays.
\section{Observations and Data analysis}
Out of the seven years of RHESSI observations since launch, an event was selected that is uniquely suited to this study. It was not only observed by RHESSI during the impulsive phase, but also by the Phoenix-2 spectrometer \citep{Mes99}  and the Nan\c{c}ay radioheliograph \citep{Ke97}. Additional observations from the GOES/SXI X-ray instrument \citep{Hi05} are available. The event was a non-occulted limb event, thus the geometry could be studied with only minor projection effects. 

The selected event occurred on August 23rd 2005. The hard X-ray peak-time (in the 25-50 keV energy band as observed by RHESSI) was at 14:37 UT. The GOES class M3 event allows the study of the entire loop geometry including the hard X-ray coronal source and footpoints.  The time evolution of the event is shown in Fig.~\ref{spectrogram}. The top panel of the figure displays the 80 to 1000 MHz part of the Phoenix-2 spectrogram. There were two extended time intervals of spike activity, lasting from 14:35:30 to 14:43:15 UT and from 14:43:25 to 14:45:00 UT. The first interval shows one (and possibly two) harmonic structures in the typical 1:1.4 ratio \citep{Be87}, and the second group is randomly distributed. In the bottom panel of Fig.~\ref{spectrogram}, the time evolution of the flare is illustrated by light-curves in hard X-rays (RHESSI) and soft X-rays (GOES). The flare displays several hard X-ray peaks starting at 14:22. The spikes first appeared at the onset of the main hard X-ray peak (the most intense peak in the 50-100 keV band).
\begin{figure*}[ht!]
\centering
\includegraphics[width=15.5cm]{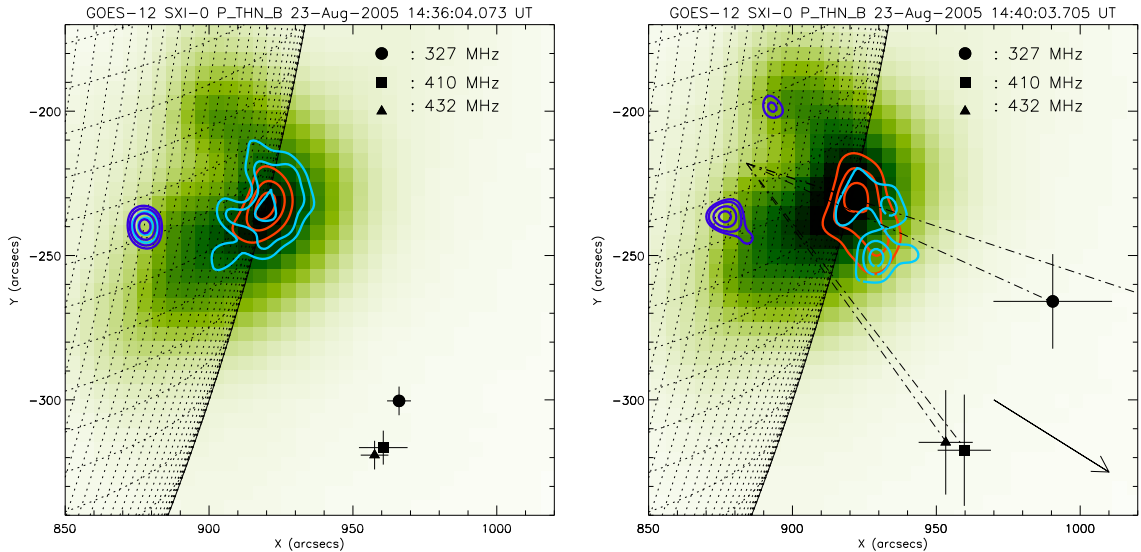}
\caption {GOES/SXI images taken during time intervals of the spike activity. RHESSI contours of the hard X-ray coronal sources are shown in light gray (light blue in the color version) at 18-22 keV. The RHESSI observations at 6-12 keV (dark gray (red in the color version), thermal) and 25-50 keV (black, footpoints) are also overlaid. The time-averaged centroid positions of the radio sources are indicated with different symbols for the different frequencies. The error bars indicate the standard deviation of the individual source positions. The arrow points in the direction of the CME as seen from Sun center. } 
\label{positions}
\end{figure*}
\subsection{X-ray observations}
The flare was observed by both RHESSI and GOES/SXI during the impulsive phase. GOES images provide information about the thermal plasma at a few MK. 
RHESSI observations were used to image the soft and hard X-ray component of the coronal source as well as the footpoints, and to determine their positions. CLEAN images were generated using detectors 3-8. Figure \ref{positions} displays the contours of images taken from 14:36:00-14:38:00 and from 14:43:15-14:45:00. The thermal component of the coronal source was imaged at energies in the range 6-12 keV, and the footpoints at 25-50 keV. The coronal source could be imaged up to energies from 18-22 keV. At higher energies, the footpoints become dominant. The coronal source spectrum as found from imaging spectroscopy indicates a nonthermal component at higher energies, where the transition between the thermal and nonthermal emission is found to occur at around 20 keV. Thus, the emission in the 18-22 keV energy band is expected to be at least partly of nonthermal origin.
\subsection{Radio observations}
Spatial information about the observed spikes was extracted from Nan\c{c}ay data at center frequencies of 327.0, 410.5 and 432.0 MHz and timesteps of 150 ms. The averaged beam widths are 170$''$, 132$''$, and 126$''$, respectively. We determined the centroid positions of the sources at each frequency by fitting a two-dimensional Gaussian to the images. As the signal-to-noise ratio in the spikes exceeds 100, the centroid position can be determined to an accuracy of a few arcseconds. The absolute accuracy is limited by instrumental and ionospheric effects to $\pm$20$''$, as comparisons with the VLA have shown.  We define the spike location for each frequency as the time-averaged position of the individual positions, using the standard deviation as a measure of the uncertainty. We note that the error in the mean is considerably smaller, making the positional difference between the 327 MHz source from the other two sources statistically significant. However, the displacement of the 327 MHz is lateral, not vertical as in a standard atmosphere.

The radio source at each of the Nan\c{c}ay frequencies can be divided into a continuous background component and a component associated with enhanced emissivity attributed to spikes (see Fig.~\ref{huetli}). To differentiate between the two components, the radio flux at each frequency was averaged over the observed time-interval. Positions associated with a radio flux higher than this average flux were allocated to the spike emission and used when computing the spike location in Fig. \ref{positions}. 
\section{Results}
In the presented event, spikes first appear at the onset of the main hard X-ray peak and continue on through the decay phase of this peak and the onset of another, less intense hard X-ray peak.  Although the hard X-ray emission reaches a local minimum when there is least spike activity, no distinct temporal correlation between X-ray flux and spikes can be identified.
\subsection{Spatial relation}
The main results are shown in Fig.~\ref{positions}. Inside the GOES image in the figure, a soft X-ray loop can clearly be seen. At the bottom of this loop, RHESSI footpoints are indicated by 40, 50, 70, and 90 percent contours. The northern footpoint is weaker by an order of magnitude than the southern footpoint and can only be seen in the second time interval. The thermal and nonthermal components of the coronal source are spatially coincident and situated at the top of the soft X-ray loop outlined by GOES.
\subsubsection{Flare geometry and source positions}
\begin{figure*}[th!]
\includegraphics[width=16cm, height=4.7cm]{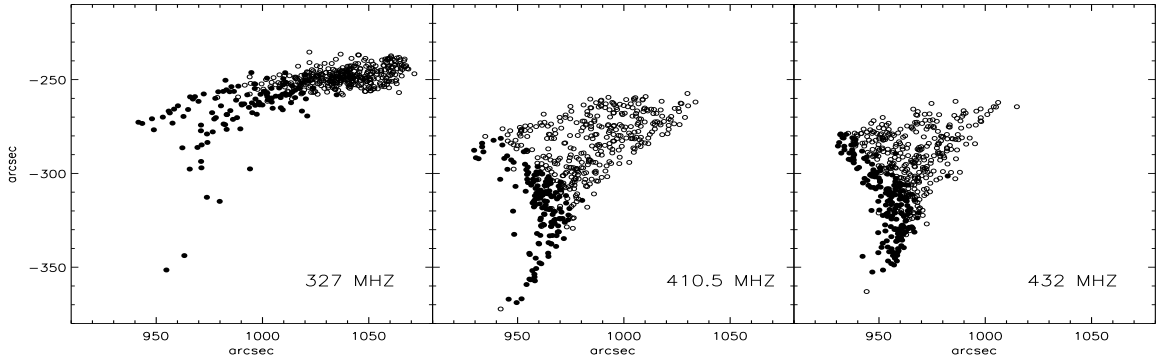}
\caption {Individual centroid positions of the radio sources during the time interval of 14:43:15-14:45:00. \textit{Dark dots} indicate locations associated with a radio flux higher than the time-averaged flux, \textit{open circles} indicate locations associated with a radio flux lower than the time-averaged flux.}
\label{huetli}
\end{figure*}
The radio sources are partially resolved, i.e., they are both slightly larger than the beam. At all frequencies, the average centroids are clearly displaced from the flaring site in SXR and HXR. The displacement of the radio sources from the coronal hard X-ray source ranges from 79\arcsec\ (source at 327 MHz during time interval 1) to 97\arcsec\ (410 MHz source during time interval 2). The distance of the radio sources from the footpoints ranges from 108\arcsec\ (southern footpoint from 327~MHz source during time interval 1) to 137\arcsec\ (northern footpoint from 410~MHz source during time interval 2). The projected height of the radio sources above the limb ranges from 38 800 km (432~MHz source during time interval 2) to 54 700~km (327~MHz source during time interval 2). 

The displacement of the radio sources is even more striking in terms of the geometry of the flare loop. Using the centroid positions of the coronal X-ray source and the footpoints, the loop is axially symmetric with respect to the perpendicular from the coronal source to the line connecting the footpoints (top dash-dotted line in Fig. \ref{positions}, right). We computed the angle between this perpendicular and the line connecting the middle between the two footpoints and the spike sources (dot-dashed lines in Fig. \ref{positions}, right). The spike sources are offset by an average angle of 28\degr\ from the perpendicular of the coronal source. The smallest offset is found for the 327 MHz source during the second time interval (8\degr), and the largest offset is 38\degr, found at 432 MHz during the second time interval. 

We may add here that a CME was later observed by SOHO/LASCO.
The CME propagates at a radial angle of about 32\degr\ relative to the solar equator. The radio sources, seen from the Sun center at 15 - 19\degr\ relative to the equator, are thus in a region between the main flare site and the estimated propagation path of the CME. The propagational direction of the CME is indicated by an arrow in Fig. \ref{positions}. The observed onset is far outside of the field of view shown in the figure. These observations do not prove any direct connection between the CME and the radio sources. However, they provide further evidence that activity is taking place in a large volume around the flaring site.

An interesting asymmetry is revealed in Fig.~\ref{huetli} which shows the individual, fitted spike source locations. For each frequency, the centroid positions from the Gaussian fits are plotted for all 500 time bins of the second interval. The first time interval shows a similar pattern. The individual positions of the spikes are scattered over about 100\arcsec\ in the x and y directions over time, exceeding the positional inaccuracy.

\section{Discussion and conclusions}
The radio sources emitting decimetric spikes have been found to be displaced significantly and to have an altitude at least twice as high as the coronal X-ray source. Spikes at different frequencies do not appear at the same position (Fig. \ref{positions}). However, the observed position of solar radio sources may deviate from the original position because of coronal effects. Coronal scattering was proposed to produce the observed source size of typically between 40\arcsec\ and 90\arcsec\ at 333 MHz \citep{Zl92}. Figure \ref{huetli} suggests that the position differences are not caused by an entirely random process. 

Refraction in the corona moves the apparent position inward. Thus, the true source position could be at an even higher altitude than found from the presented observations. Therefore, the observed displacement of the radio sources from the X-ray sources is real.
The spike positions determined here indicate that the plasma above the hard X-ray source is highly active and contains nonthermal electrons emitting coherent radio waves. It cannot be excluded that those same electrons are also producing X-rays that are not observed in the presented event because of the emission intensity from the main flaring site and RHESSI's limited dynamic range. Observational evidence of high coronal X-ray emission was found by e.g., \citet{Hu01} and \citet{Kr07} in highly occulted events.  Coronal X-ray sources moving outward during the course of the flare were observed by eg. \citet{Ga02} and \citet{Sui03}, suggesting a shift in the energy release site occurred during the course of the flare.

The spike observations indicate the presence of non-thermal electrons, possibly originating from additional, spatially 
separated energy release in the high corona. Considering that the spikes occur during the impulsive phase, there 
appears to be a causal link, direct or indirect, between the spikes and the main flare energy release near or inside the 
coronal source.

\mdseries
\begin{acknowledgements}
The authors wish to thank Nicole Vilmer for providing the Nan\c{c}ay data. RHESSI data analysis at ETH Zurich is supported by ETH grant TH-1/04-2. The construction and operation of Phoenix-2 is credited largely to Christian Monstein and is partially funded by the Swiss National Science Foundation Grant 200020-121676. This research made use of the HESSI European Data Center (HEDC) at ETH Zurich and NASA's Astrophysics Data System Bibliographic Services.
\end{acknowledgements}

\bibliographystyle{aa}
\bibliography{mybib}

\end{document}